# EFFECT OF MICROSCALE TURBULENT STRUCTURES DYNAMICS ON FORCED CONVECTION IN TURBULENT POROUS MEDIA FLOW

**Ching-Wei Huang[1], Vishal Srikanth[1], Andrey V. Kuznetsov[1*]**

[1]Department of Mechanical and Aerospace Engineering, North Carolina State University, Raleigh, NC 27695, USA

## ABSTRACT

The influence of microscale flow structures (smaller than the pore size) on turbulent heat transfer in porous media has not been yet investigated. The goal of this study is to determine the influence of the micro-vortices on convection heat transfer in turbulent porous media flow. Turbulent flow in a homogeneous porous medium was investigated using Large Eddy Simulation (LES) at a Reynolds number of 300. We observed that the convection heat transfer characteristics are dependent on whether the micro-vortices are attached or detached from the surface of the obstacle. There is a spectral correlation between the Nusselt number and the pressure instabilities due to vortex shedding. A secondary flow instability occurs due to high pressure regions forming periodically near the converging pathway between obstacles. This causes local adverse pressure gradient, affecting the flow velocity and convection heat transfer. This study has been performed for obstacles with shapes of square and circular cylinders at porosities of 0.50 and 0.87. Understanding the dominant modes that affect convection heat transfer can aid in finding an optimum geometry for the porous medium.

## 1. INTRODUCTION

Models of turbulent porous media flow are useful for the systematic study for applications such as canopy flows, pebble bed nuclear reactors, heat exchangers, porous chemical reactors, and crude oil extraction [1]. Macroscopic turbulence models for porous media flows have been developed throughout the years [2–5], combining the Volume Average Theory (VAT) [6] with Reynolds Averaging (RA). Reynolds Averaged Navier-Stokes (RANS) simulations of microscale porous media flow have been used to determine coefficients for the RA-VAT models [7–9]. However, the results from microscale RANS simulations are constrained by the modelling error [10].

Jin et al. [11] and Uth et al. [12] performed Direct Numerical Simulation (DNS) of forced convection flows in porous media, which suggested that the pore size of the porous medium determines the maximum size of turbulent eddies. The study led to the development of a mixing-length macroscale model based on the mixing layer hypothesis by Jin & Kuznetsov [13]. DNS studies by He et al. [14] verified that the turbulence integral length scale is ~10% of the obstacle diameter in a closely packed porous medium. Turbulent energy transport is crucial for studying convection heat transfer in porous media. Several macroscopic energy models have been developed that make use of the gradient diffusion hypothesis in conjunction with the assumption of thermal equilibrium between the solid and fluid phases [9],[15].

Microscale studies show that the heat transfer efficiency between the obstacle surface and the fluid increases with an increase in the Reynolds number and obstacle diameter [16,17]. The microscale simulations for square rods [15], circular rods [18], and elliptic rods [19] revealed that the thermal dispersion varies drastically with the obstacle shape. The functional dependence of the Nusselt number on porosity changes with the obstacle shape [20]. High Resolution LES studies of finite pebble beds show that hot spots appear on the surface of the pebbles are highly unsteady, in which their locations move over time [21], highlighting the importance of a transient analysis. The microscale distribution of the Nusselt number from DNS studies [16] shows that the unsteady wake region contributes the least towards heat transfer. Turbulent thermal mixing for circular rods increases with an increase in the Reynolds number and approaches an asymptotic value at higher Reynolds numbers [22].

*Corresponding Author: avkuznet@ncsu.edu

## 2. METHODS

Cylindrical solid obstacles have been arranged in a simple square lattice to represent a homogeneous, anisotropic porous medium. A three-dimensional simulation domain (see Fig. 1) has been constructed using 4 unit cells along the *x*- and *y*- directions, and 2 unit cells in the *z*- direction based on the inferences from Jin et al. [11] and Uth et al. [12], forming a $4s \times 4s \times 2s$ Representative Elementary Volume (REV). We use cases with $2s \times 2s \times 2s$ REV for grid study; all simulation cases are listed in Table 2. Periodic boundary conditions are used to impose an infinite span in all directions to avoid finite boundary effects. The phenomena that are observed inside the periodic domain can thus be linked to the porous medium alone. Two values of porosity ($\varphi$) are studied, 0.50 and 0.87, while the Pore Scale Reynolds number ($Re_p$) is 300 unless specified otherwise. The $Re_p$ was maintained using a constant applied pressure gradient ($dp/dx$) as the driving force. These parameters are defined as

$$\varphi = 1 - \frac{\pi}{4}\left(\frac{d}{s}\right)^2 \tag{1}$$

$$Re_p = \varphi \cdot \frac{u_m d}{\nu} \tag{2}$$

where $d$ is the diameter of cylindrical obstacles, $u_m$ is the mean velocity in the *x*-direction, and $\nu$ is the kinematic viscosity of the fluid. The temperature of the walls of the cylindrical obstacles was set to a constant value of 353K, while the average temperature of the inlet flow was set to 323K. With this boundary condition setting, the characteristic temperature difference, $\Delta T$, was 30K. The Prandtl number ($Pr$) was kept constant at 6.99. We use RANS and LES methods to simulate the microscopic flow field. The Dynamic One-equation TKE (DOTKE) subgrid model has been used for LES, and the Realizable *k*-*ε* model has been used for RANS. Simulations were performed using the commercial CFD code ANSYS Fluent 16.0.

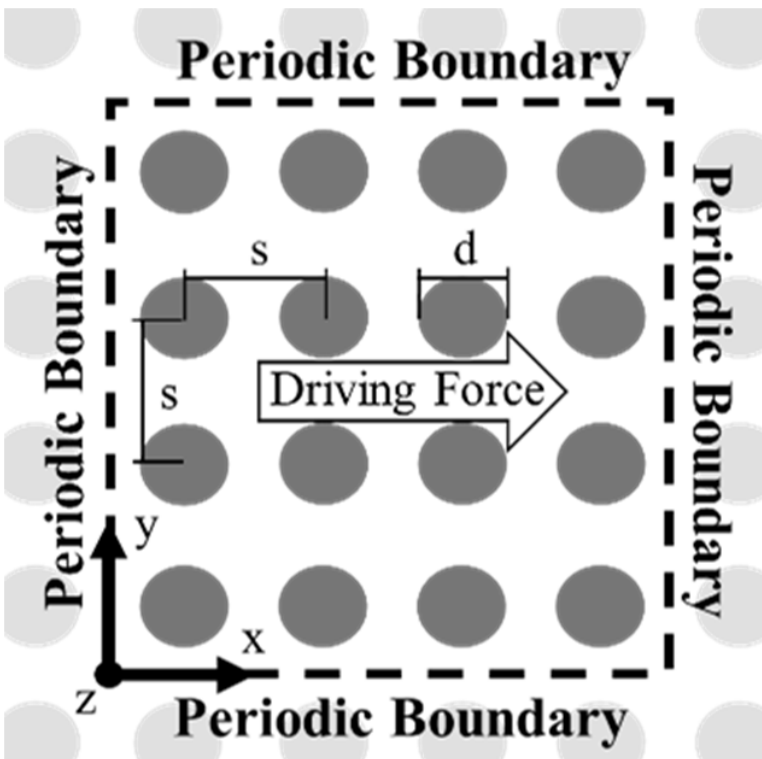

**Fig. 1** The Representative Elementary Volume (REV) geometry of a porous medium. The distance between centers of the obstacles *s* and the obstacle diameter *d* are also shown in the figure.

Grid resolutions for LES cases as well as maximum values of non-dimensional near-wall grid spacing, $\Delta y^+_{max}$, are shown in Table 2. The LES Index of Quality (LES_IQ) [23] is used to provide the fraction of the total turbulence kinetic energy that is resolved by the grid. Assuming recommended 80% of the energy resolved in LES [24], the simulated LES_IQ should be greater than 0.8. Remarks from Celik *et al.*[23] indicate that simulations may be considered to be of DNS quality with LES_IQ > 0.9. The volume averaged LES_IQ for all the LES simulations in this work meet this criterion after time averaging. The minimum and spatially averaged values of LES_IQ at an instant in time are reported in Table 3. For the LES cases, the turbulence kinetic energy spectrum is used to identify the scale regimes of turbulence that have been resolved in this work. The turbulence kinetic energy spectra ($E_{ii}/3$) versus the wavenumber ($k \cdot s$) for the LES test cases are shown in Fig. 2.

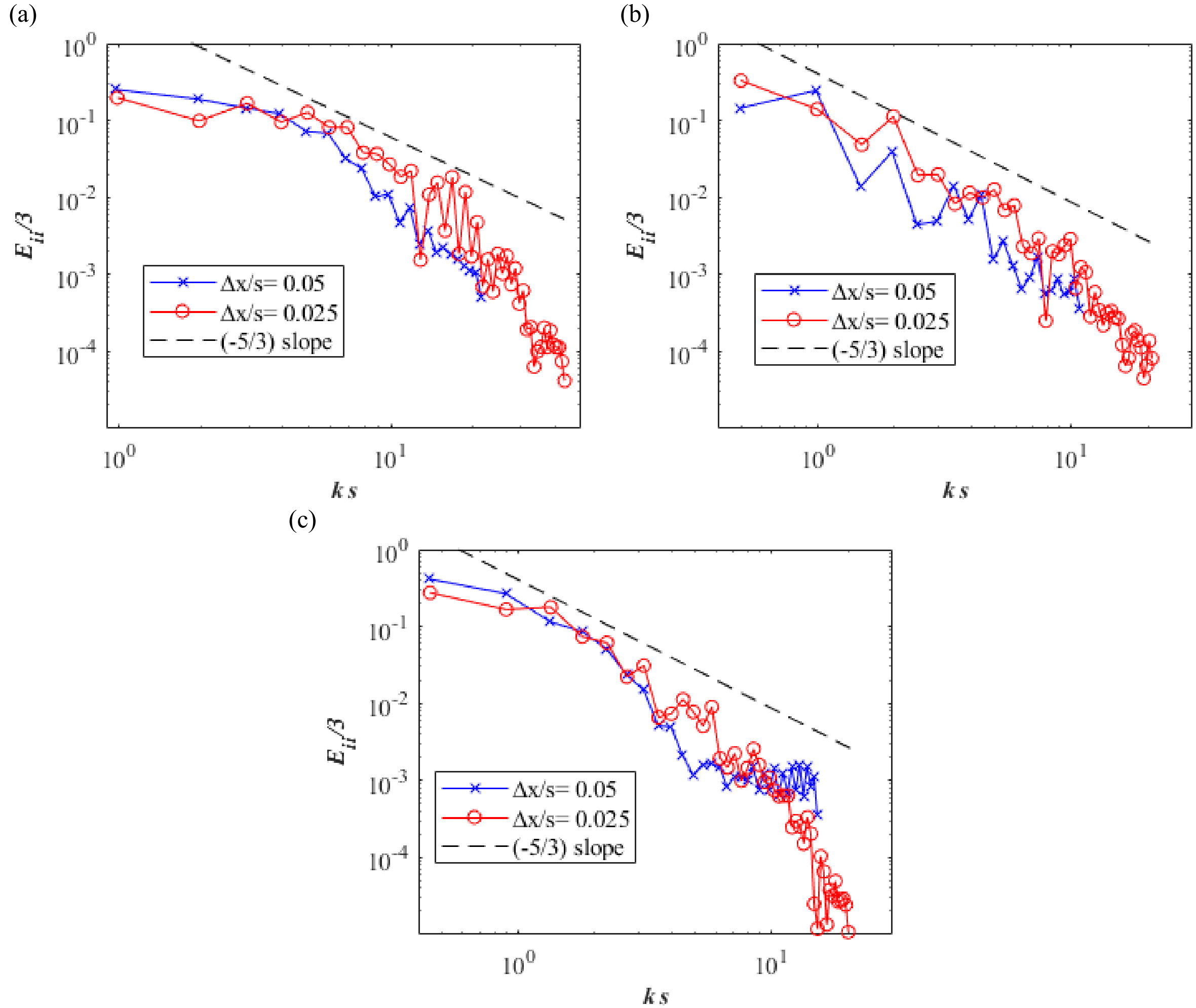

**Fig. 2** Turbulence energy spectrum for LES cases (a) $\varphi = 0.50$ with circular obstacles; (b) $\varphi = 0.87$ with circular obstacles; (c) $\varphi = 0.87$ with square obstacles. The dashed line corresponds to the -5/3 slope on the log plot and $\Delta x$ is maximum grid spacing.

Heat transfer in porous media is closely linked to the formation, propagation, and dissipation of flow structures. The use of RANS simulations to determine coefficients for the RA-VAT models will not accurately capture time dependent effects such as vortex shedding or Hopf bifurcations in periodic porous media, which leads to a considerable error when using them. This is demonstrated from the difference between the mean Nusselt number ($Nu_m$) of the RANS and LES simulations shown in Table 4.

**Table 2** The maximum value of non-dimensional near-wall grid spacing, $\Delta y^{+}_{max}$, measured on the surface of the solid obstacles for the grid resolution test cases. These are small areas with high $\Delta y^{+}$ values, overall $\Delta y^{+}$ values on the obstacle surfaces are kept below 1.

| $\Delta y^{+}_{max}$ | | |
|---|---|---|
| Case | Coarse grid, $\Delta x_{max}/s$= 0.05 | Intermediate grid, $\Delta x_{max}/s$= 0.025 |
| $\varphi = 0.50$ | 1.06 | 1.16 |
| $\varphi = 0.87$ | 1.68 | 1.81 |
| $\varphi = 0.87$(square) | 1.87 | 1.65 |

**Table 3** The value of LES_IQ measured in the fluid volume for the grid resolution test cases. Both the minimum and the volume-averaged values are reported (ranges from 0 to 1, high values indicate high resolution with a large fraction of the turbulence kinetic energy being resolved).

| **LES_IQ** | | | |
|---|---|---|---|
| Porosity φ | | Coarse grid, $\Delta x_{max}/s$= 0.05 | Intermediate grid, $\Delta x_{max}/s$= 0.025 |
| 0.50 | minimum | 0.22 | 0.44 |
| | average | 0.81 | 0.95 |
| 0.87 | minimum | 0.70 | 0.73 |
| | average | 0.96 | 0.98 |
| 0.87(square) | minimum | 0.69 | 0.74 |
| | average | 0.95 | 0.98 |

**Table 4** Comparison of RANS and time averaged LES results at $\varphi = 0.50$ and $Re_p = 500$. For mean *x*-velocity ($u_m$) and mean Nusselt number ($Nu_m$).

| | RANS | LES(time averaged) |
|---|---|---|
| $u_m$ | 0.1258 | 0.1296 |
| $Nu_m$ | 59.953 | 72.650 |

## 3. RESULTS AND DISCUSSION

For $Re_p = 300$, the von Kármán instability is observed for the flow around each obstacle. The amplitude of lift oscillation from the von Kármán instability depends on the obstacle shapes as well as porosity ($\varphi$). A symmetry-breaking flow bifurcation is expected to occur at this $Re_p$. The mean flow direction of the porous medium flow can either coincide with the driving force or deviate from the direction of the driving force. The phenomenon originates from asymmetrical vortex breakdown and it is highly sensitive to the geometry of the porous medium. The formation of flow deviation has been studied in detail by Srikanth *et al*. [25]. The deviating direction around the flow of each cylinder is synced for obstacles in the same column, where any sway from the sync will be corrected by the neighboring obstacle flow of the same column. Fig. 3 illustrates the lift coefficient ($C_l$) over time for the case of $Re_p = 300$, $\varphi = 0.87$ with circular cylinder obstacles (case A2). Only the 1st column of obstacles in the REV are plotted to avoid the figure becoming too busy. The direction of flow deviation can be observed from the $C_l$ oscillation over time. Fig. 4(b) illustrates a snapshot of this case, where $C_l$ is at a positive value for the 1st column of obstacles. The existence of this phenomenon highlights the importance of micro-vortices in transport in porous media, since it is a source of enhanced flow mixing. Different geometries will have different flow features due to the interaction of the micro-vortices with the surfaces of the obstacles forming the porous medium.

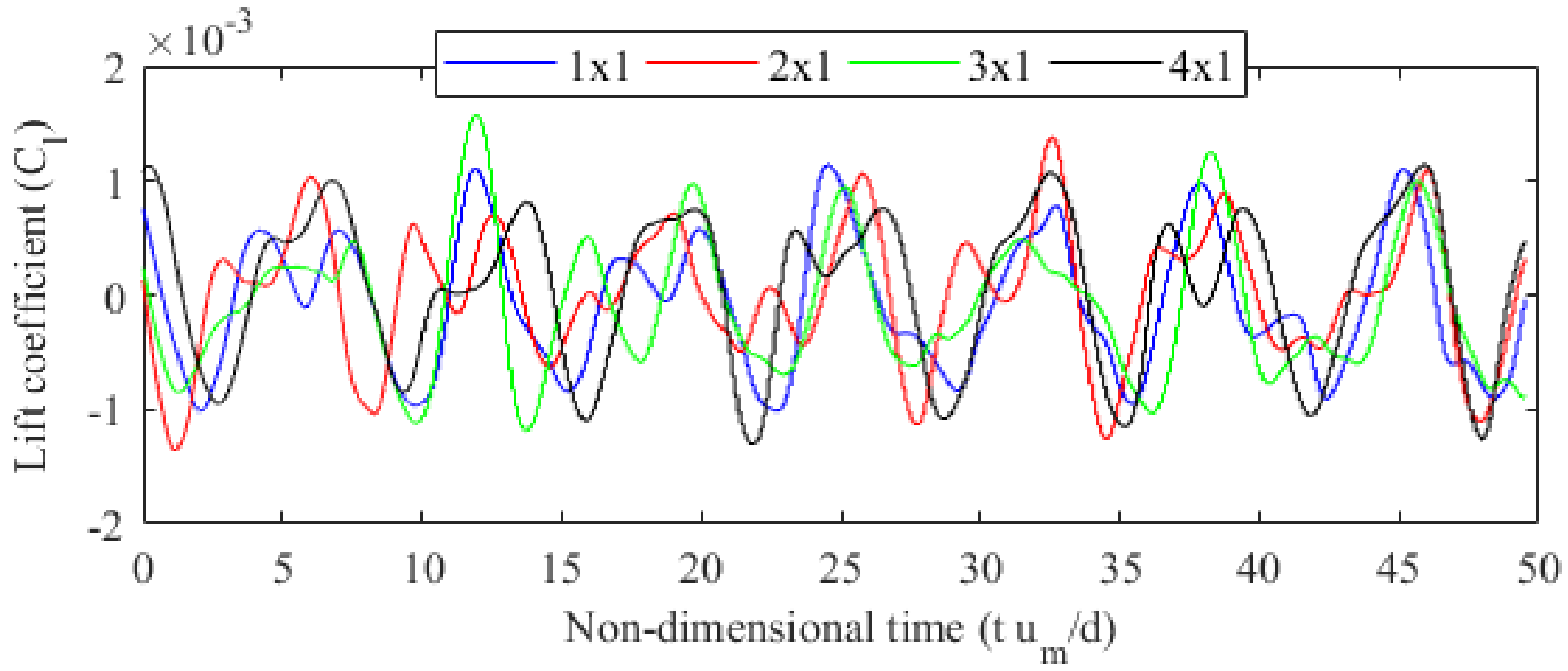

**Fig. 3** Change of lift coefficient ($C_l$) over time for obstacles in the 1st column for case A2 ($Re_p = 300$, $\varphi = 0.87$ with circular cylinder obstacles).

(a) Case A1

(b) Case A2

(c) Case A3

Temperature (K): 323 329 335 341 347 353

| Case ID | Obstacle Shape | Porosity | Type |
|---|---|---|---|
| A1 | Circular | 0.50 | LES |
| A2 | Circular | 0.87 | LES |
| A3 | Square | 0.87 | LES |
| B1 | Circular | 0.50 | RANS |

**Fig. 4** Streamlines and temperature distribution in a porous medium consisting cylindrical obstacles with porosity of (a) Case A1, $\varphi = 0.50$ with circular obstacles; (b) Case A2, $\varphi = 0.87$ with circular obstacles; (c) Case A3, $\varphi = 0.87$ with square obstacles.

The micro-vortices that are generated behind the solid obstacle are pockets of slow-moving fluid, which insulate the solid obstacle from the fast-moving fluid. This reduces the effective surface area that is available for heat transfer. For the circular cylinder obstacles, a more diffuse temperature distribution is observed in a shedding vortex system. A shedding vortex system will perform the additional role of carrying away pockets of heat from the obstacle surface. If the amplitude from the von Kármán instability is not sufficient for the wake vortices to break into the primary duct flow, a recirculating vortex system is formed between two neighboring obstacles in the same row. This is illustrated in case A1 (Fig. 4(a) and Fig. 5). The recirculating vortex system traps heat in the streamwise pore space rendering the vortex-covered portion of the surface area less conducive for heat removal. On the other hand, if the wake vortices can break into the primary flow, vortices will periodically form near the obstacle surface and carry heat when dissipated into the primary flow, as illustrated in case A2 (Fig. 4(b)). The temperature distribution in this case will have steep gradients, which increases the surface heat transfer rate. A similar behavior can be found in case A3 (Fig. 4(c)), where the obstacles are square cylinders. The wake vortices are able to break into the primary flow, but due to the sharp corners of the square obstacle shape, the location of the separation points does not change over time. This results in a smaller amplitude of intensity when vortices periodically form and dissipate, as well as a constant contact area between the wake vortices and the obstacle surface.

The difference in terms of vortex shedding here is that in case A1 the vortices formed are shed into the recirculating vortices, while in case A2 and A3 the vortices are shed into the primary flow. The space between the obstacles (porosity) is the cause for this difference in vortex shedding. Two distinct momentum transport processes, micro-vortex transport from the surface and turbulent mixing in the bulk flow, govern heat transfer inside porous media. The dominance of each process is observed to be sensitive to the geometry of the porous medium and the Reynolds number of the flow.

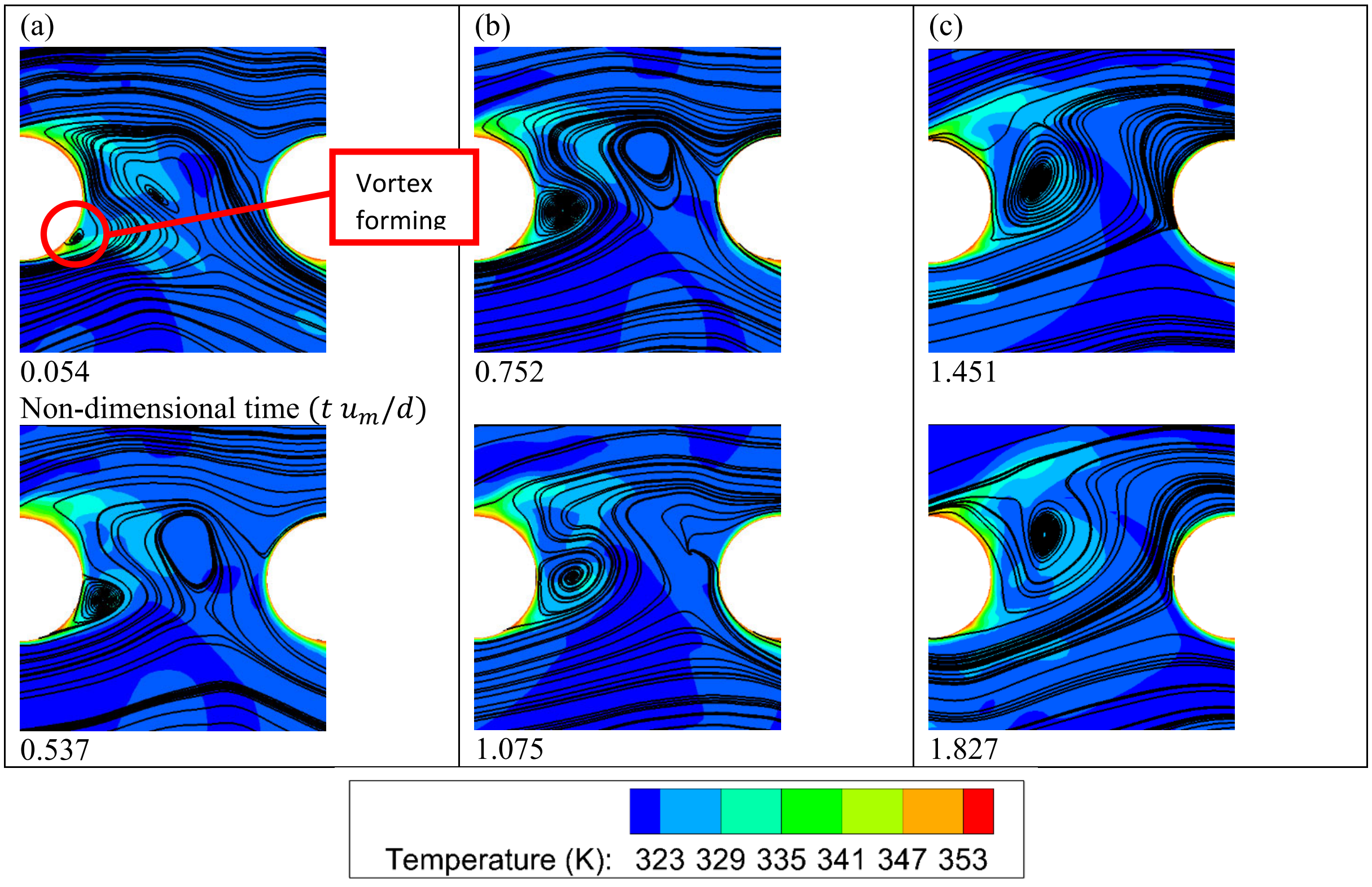

**Fig. 5** Example of vortex shedding and its contribution to heat transfer from case A2 (a) Vortex beginning to form while the previous vortex is shed and dissipated downstream. (b) A developed vortex that acts as an insulation on the obstacle surface. (c) Vortex is shed downstream, a new vortex starts forming.

From our results, the cause of the difference between the RANS and LES results shown in Table 4 comes from dynamic behavior of micro-vortices. Convection heat transfer is mainly affected by the following behaviors of the micro-vortices: vortex recirculation, vortex shedding and flow stagnation. Of these three, flow stagnation will also affect the intensity of vortex shedding in addition to its own effect on heat transfer. The recirculating vortices mainly act as an "insulation" region when in contact with the obstacle surface, with a lower convection heat transfer in this region. This is due to a lower velocity and smaller temperature gradient in the recirculating vortices, as shown in Fig. 4. They are attached to the wake region of the porous medium obstacles. This effect is captured by the RANS results, as the recirculating vortices do not change significantly over time.

The shedding micro-vortices transport heat away from the surface and dissipate it in the primary flow. The vortex shedding process starts by vortex formation on the obstacle surface. Because of the higher temperature gradient and mixing from the newly formed vortex, the surface in contact with the vortex has a higher heat transfer rate at this stage. After the vortex is formed, it grows while being attached to the surface. The vortex acts as an insulation on the obstacle surface in this stage, similar to that of a recirculating vortex. When the vortex is too large to maintain stability on the obstacle surface, it is shed downstream, and a new vortex starts forming. This process is shown in Fig. 5.

Flow stagnation happens due to high pressure regions forming periodically near the converging pathway between two obstacles. This is caused by stagnation pressure oscillation caused by competition between the pressure and unsteady forces, as shown in Fig. 6. The vortices that shed periodically from the obstacle surface are similar to that of a von Karman vortex, which introduces oscillations to the forces acting on the obstacle surface. The intensity of the vortex shedding is monitored using the drag force (Fig. 6), which consists of viscous and pressure drag components. The pressure drag has a higher magnitude and is therefore more dominant than the viscous drag. The higher frequency oscillations of the pressure force represent vortex shedding on the obstacle surface. This is verified by matching the vortex shedding visualization in the animation to the oscillation of the pressure force in Fig. 6.

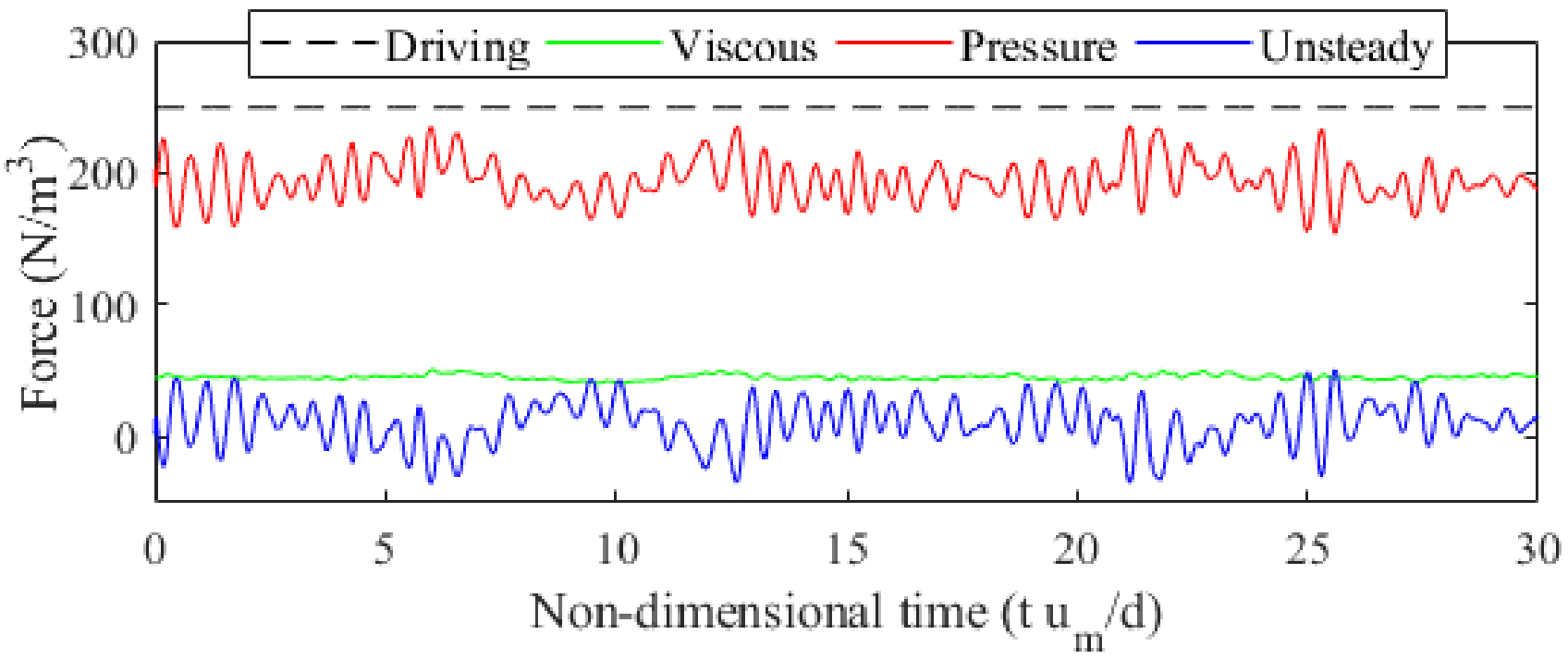

**Fig. 6** Momentum balance of $x$-direction forces within REV for porous media flow with circular cylinders, $\varphi = 0.50$ and $Re_p = 300$ (case A1).

The flow stagnation affects the intensity of vortex shedding, which can be seen in the change of the pressure force oscillation amplitude over time. To observe the correlation between heat transfer and the effects of flow stagnation and vortex shedding, the surface heat transfer rate is represented by the Mean Nusselt Number ($\mathrm{Nu_m}$); the effects of flow stagnation and vortex shedding, which can be seen in the change in pressure force, is represented by the Drag Coefficient ($\mathrm{C_d}$). To separate the vortex shedding and flow stagnation effects that are in the force signal, the FFT of the signals was computed (Fig. 7). The amplitude of both the $\mathrm{C_d}$ and $\mathrm{Nu_m}$ signals are converted to power (dB), as shown in equation 3, for comparison.

$$Power(\mathrm{P}) = 10 * log_{10}(\frac{P}{P_m}) \ \ \mathrm{dB} \tag{3}$$

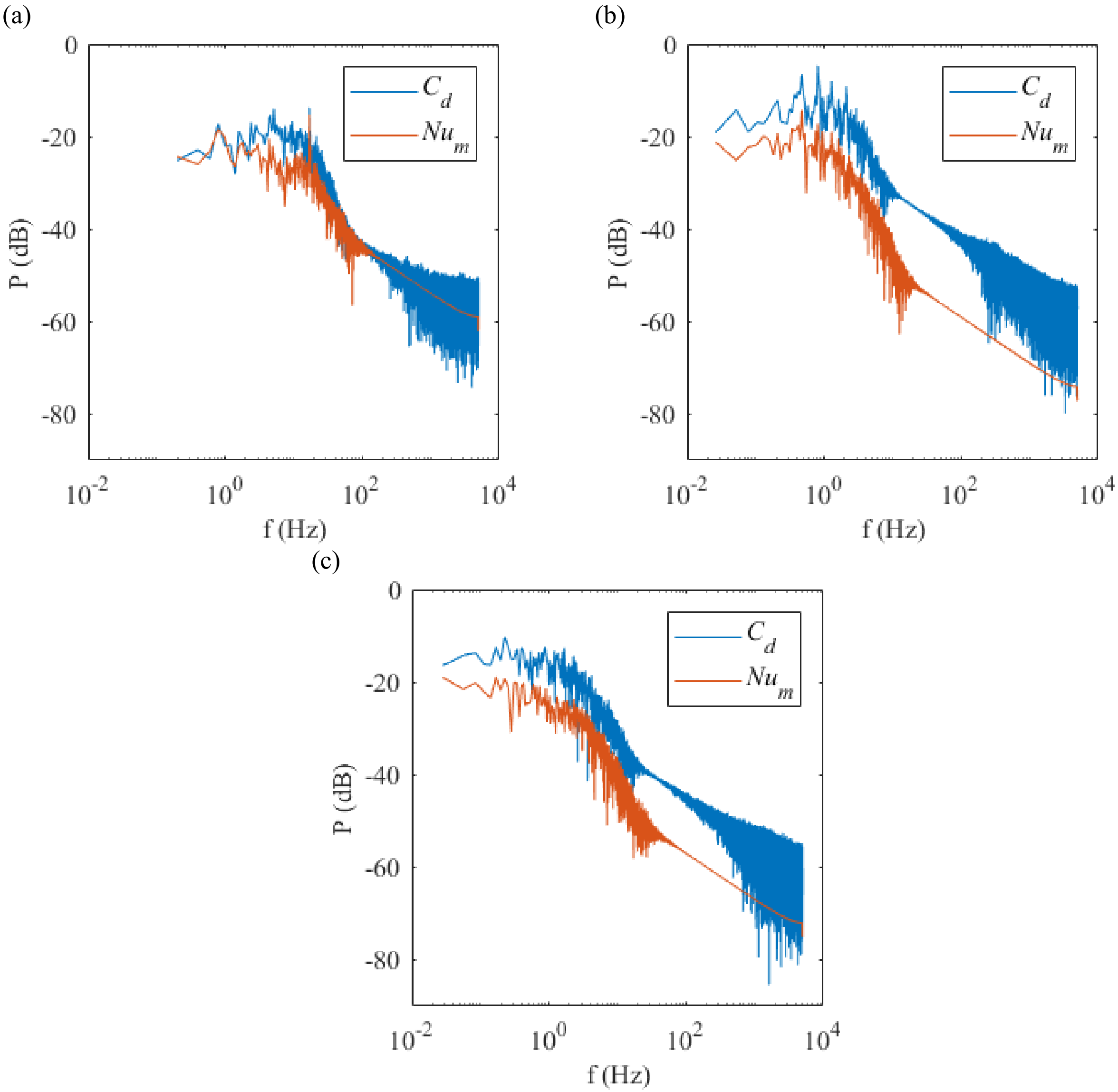

**Fig. 7** Fast Fourier Transform (FFT) of the Drag Coefficient ($\mathrm{C_d}$) and Mean Nusselt Number ($\mathrm{Nu_m}$) for (a) Case A1, $\varphi = 0.50$ with circular obstacles; (b) Case A2, $\varphi = 0.87$ with circular obstacles; (c) Case A3, $\varphi = 0.87$ with square obstacles.

The FFT of the drag coefficient at $\varphi = 0.50$ reveals a lower frequency peak that represents flow stagnation, then a spectrum of frequencies representing the vortex shedding. Comparing $\mathrm{C_d}$ and $Nu_m$ in case A1 (Fig. 7(a)) we can see a correlation between the two signals, which provides evidence that the dynamics of heat transfer is caused by the effects of vortex shedding and flow stagnation. The case A2 (Fig. 7(b)) shows no dominant frequency, but rather a spectrum of frequencies. The lower frequency represents the fact that flow stagnation is still present, but vortex shedding is no longer constrained to a predominant frequency. The change in frequency distribution occurs because the vortices are given more space to develop due to the increase of distance between the obstacle walls. The space allows for the vortex shedding to interact and propagate with the primary flow, resulting with a wider range of frequencies in the change of $C_d$. Case A3 shows similar traits to A2, but with a smaller amplitude in power. Compared to circular obstacles, the sharp corners of the square obstacle shape restrict the location of the separation points to change over time. This results in a smaller amplitude of intensity when vortices periodically form and dissipate, which shows as smaller power amplitudes (P in Fig. 7) in the FFT of $C_d$ and $\mathrm{Nu_m}$. The combined effect of vortex shedding, and flow stagnation is missing from the steady state RANS result and is captured in the LES simulation, causing ~17% difference between the $Nu_m$ values predicted by these simulations. The signals diverge above the frequency of ~$10^2$, which indicates there is no direct relation between heat transfer and flow at these higher frequencies (smaller eddies do not impact surface $Nu_m$).

## 4. CONCLUSIONS

Micro-vortices play an essential role in porous media flow, and by extension, in convection heat transfer in porous media. The FFT of $Nu_m$ and $C_d$ signals show a match in dominant frequencies, which is supported by flow visualization. This confirms our hypothesis that vortex shedding and flow stagnation are characterized by the time dependent dynamics that predominantly affects heat transfer.

The occurrence of vortex shedding and flow stagnation are sensitive to the porosity and obstacle shapes of the porous medium. For a porous medium composed of circular cylinders, in the case A1 ($\varphi = 0.50$ with circular obstacles), both vortex shedding and flow stagnation occur. When porosity is increased to $\varphi = 0.87$ (case A2), vortex shedding is characterized by a spectrum of frequencies and there is no longer a dominant frequency. It is likely that the change in frequency distribution is because flow in porous media with higher porosity behaves closer to that of open flow over an isolated obstacle. When changing the obstacle shape to square cylinders, case A3 ($\varphi = 0.87$ with square obstacles), the sharp corners of the square obstacle restrict the location of the separation points. This results in a frequency distribution that is similar to case A2, but with a smaller amplitude in power due to the restriction from the obstacle shape.

Further investigations are needed to establish the role of time dependent dynamics of vortex shedding and flow stagnation on heat transfer at higher Reynolds numbers.

## ACKNOWLEDGMENT

AVK acknowledges with gratitude the support of the National Science Foundation (award CBET-2042834), the Alexander von Humboldt Foundation through the Humboldt Research Award, and the Extreme Science and Engineering Discovery Environment (XSEDE), which is supported by National Science Foundation grant number ACI-1548562.

## NOMENCLATURE

| | | | | | |
|---|---|---|---|---|---|
| $Re_p$ | Pore Scale Reynolds number | ( - ) | $\varphi$ | porosity | ( - ) |
| $Nu_m$ | mean Nusselt number | ( - ) | $Pr$ | Prandtl number | ( - ) |
| $u_m$ | mean $x$-velocity | (m/s) | $P$ | signal power | (dB) |